# New Universality Classes for Two-Dimensional $\sigma$-Models

Sergio Caracciolo
*Scuola Normale Superiore and INFN – Sezione di Pisa*
*Piazza dei Cavalieri*
*Pisa 56100, ITALIA*
Internet: CARACCIO@UX1SNS.SNS.IT
Bitnet: CARACCIO@IPISNSVA.BITNET
Hepnet/Decnet: 39198::CARACCIOLO

Robert G. Edwards
*Supercomputer Computations Research Institute*
*Florida State University*
*Tallahassee, FL 32306 USA*
Internet: EDWARDS@MAILER.SCRI.FSU.EDU

Andrea Pelissetto[*]
Alan D. Sokal
*Department of Physics*
*New York University*
*4 Washington Place*
*New York, NY 10003 USA*
Internet: PELISSET@MAFALDA.PHYSICS.NYU.EDU, SOKAL@ACF4.NYU.EDU

July 31, 1993

---

[*]Address until August 31, 1994. Permanent address: Dipartimento di Fisica and INFN – Sezione di Pisa, Università degli Studi di Pisa, Pisa 56100, ITALIA. Internet: PELISSET@SUNTHPI1.DIFI.UNIPI.IT; Bitnet: PELISSET@IPISNSVA.BITNET; Hepnet/Decnet: 39198::PELISSETTO.

**Abstract**

We argue that the two-dimensional $O(N)$-invariant lattice $\sigma$-model with mixed isovector/isotensor action has a one-parameter family of nontrivial continuum limits, only one of which is the continuum $\sigma$-model constructed by conventional perturbation theory. We test the proposed scenario with a high-precision Monte Carlo simulation for $N = 3, 4$ on lattices up to $512 \times 512$, using a Wolff-type embedding algorithm. The finite-size-scaling data confirm the existence of the predicted new family of continuum limits. In particular, the $RP^{N-1}$ and $N$-vector models do not lie in the same universality class.

Two-dimensional nonlinear $\sigma$-models are of direct interest in condensed-matter physics, and they are important "toy models" in elementary-particle physics by virtue of their close similarity to four-dimensional gauge theories. In this Letter we summarize analytical and Monte Carlo evidence [1] which calls into question some aspects of the conventional wisdom regarding two-dimensional $\sigma$-models and their universality classes.

The conventional wisdom [2] holds that two-dimensional lattice $\sigma$-models with a non-Abelian symmetry group are *asymptotically free* — i.e. the theory has a mass gap for all $\beta < \infty$ — and that the asymptotic behavior as $\beta \to \infty$ can be deduced from the lattice Hamiltonian simply by noting its symmetries and computing its perturbative expansion around an ordered state. The first part of this conventional wisdom has been questioned by Patrascioiu and Seiler [3]. Here we do not question the first part, but rather the second part: we argue that there are *new universality classes* which cannot be seen in conventional perturbation theory.

Consider the lattice $\sigma$-model taking values in the unit sphere $S^{N-1} \subset \mathbb{R}^N$, with Hamiltonian (= Euclidean action)

$$\mathcal{H}(\boldsymbol{\sigma}) = -\beta_V \sum_{\langle xy \rangle} \boldsymbol{\sigma}_x \cdot \boldsymbol{\sigma}_y - \frac{\beta_T}{2} \sum_{\langle xy \rangle} (\boldsymbol{\sigma}_x \cdot \boldsymbol{\sigma}_y)^2 \qquad (1)$$

(the so-called mixed isovector/isotensor model). According to the conventional wisdom, this is simply a "variant action" for the usual $O(N)$-invariant $S^{N-1}$ $\sigma$-model, and all paths in the $(\beta_V, \beta_T)$-plane satisfying $\beta \equiv \beta_V + \beta_T \to +\infty$ are claimed to give rise to the same continuum-limit theory. Here we shall argue that this is not so: we claim that in fact there is a *one-parameter family of continuum-limit theories*, labelled by a parameter $\lambda_{ren}$, only one of which ($\lambda_{ren} = +\infty$) is the continuum $\sigma$-model constructed by conventional perturbation theory.

To see this, recall first that at $\beta_T = +\infty$ the spins are forced to point along a single axis; thus, the isotensor correlation functions are identically 1, and the isovector correlation functions are those of the Ising model at inverse temperature $\beta_V$. In particular, there is a critical point at $\beta_T = +\infty$, $\beta_V = \beta_{c,Ising} = \frac{1}{2}\log(1 + \sqrt{2}) = 0.440686\ldots$.

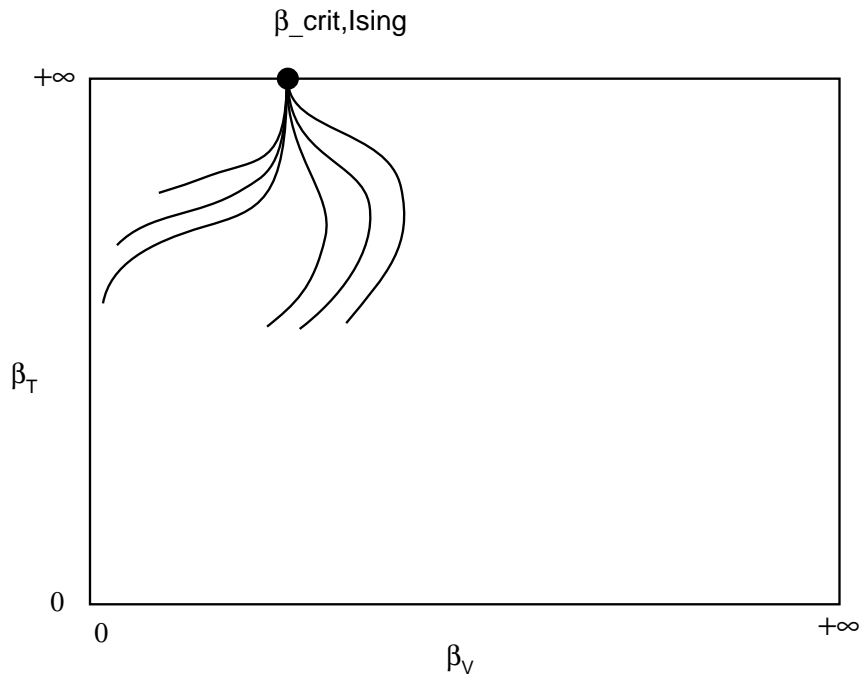

Figure 1: Proposed qualitative form of the "curves of constant physics" in the $(\beta_V, \beta_T)$-plane.

Assuming that there is no critical point in the finite $(\beta_V, \beta_T)$-plane, let us inquire about the shape of the contours of fixed isotensor/isovector mass ratio $m_T/m_V$. More generally, for any dimensionless combination $\mathcal{O}$ of *long-distance* observables, let us inquire about the shape of the contours of fixed $\mathcal{O}$; asymptotically in the continuum limit these contours are expected to be independent of the observable $\mathcal{O}$, and we call them "curves of constant physics".

We shall now argue that the curves of constant physics have the qualitative form shown in Figure 1. For concreteness, let $\mathcal{O}$ be (for now) the mass ratio $m_T/m_V$. For $\beta_V = 0$ (the $RP^{N-1}$ model), we have $m_V = \infty$ by $Z_2$ gauge invariance, hence $m_T/m_V = 0$. More generally, along any path satisfying $\beta_T \to \infty$ and $\beta_V \leq \beta_{c,Ising} - \epsilon$, we have $m_T \to 0$ while $1/m_V$ remains bounded above by its Ising value (by a rigorous correlation inequality [1]), so that $m_T/m_V \to 0$. On the other hand, it is heuristically reasonable that along any path satisfying $\beta \equiv \beta_V + \beta_T \to \infty$ and $\beta_V \geq \beta_{c,Ising} + \epsilon$, the continuum limit should be exactly that of the $N$-vector model, and in particular should have $m_T/m_V \to 2$. This is suggested qualitatively by the fact that at $\beta_T = +\infty$ and $\beta_V > \beta_{c,Ising}$, the isovector sector is spontaneously magnetized and has a finite correlation length; therefore, we expect that near such a point (at $\beta_T$ finite but large) the only possible *critical* fluctuations are spin waves around this ordered state, which are identical to those in the pure $N$-vector model. A more quantitative support for this view is provided by the expansion in powers of $1/\beta_T$ around the Ising model [4], which for $\beta_V$ fixed $> \beta_{c,Ising}$ is essentially identical to that of the pure $N$-vector model, modulo a finite renormalization of $\beta$. Putting together the two halves of the argument, we conclude that *all the curves of constant physics must approach the $\beta_T = +\infty$ axis precisely at $\beta_V = \beta_{c,Ising}$.*

It follows from this picture that there exists (at least) a *one-parameter family of continuum-limit theories*, which can be parametrized by any suitable observable $\mathcal{O}$. At one extreme is the $RP^{N-1}$ continuum theory, which is obtained along any path satisfying $\beta_T \to \infty$ with $\beta_V \leq \beta_{c,Ising} - \epsilon$. At the other extreme is the $N$-vector continuum theory, which is obtained along any path satisfying $\beta \equiv \beta_V + \beta_T \to \infty$ with $\beta_V \geq \beta_{c,Ising} + \epsilon$. Finally, there is a one-parameter family of intermediate theories, which are obtained along very special paths $\beta_T \to \infty$, $\beta_V \to \beta_{c,Ising}$ that are asymptotically tangent (to sufficiently high order) to one of the curves of constant $\mathcal{O}$.

A renormalization-group argument suggests that the curves of constant $\mathcal{O}$ have asymptotically the form
$$\beta_V = f(\beta_T) + \lambda_{ren} g(\beta_T) , \tag{2}$$
where $f(\beta_T)$ is a nonuniversal function arising from analytic corrections to scaling ("nonlinear scaling fields"), $g(\beta_T)$ is a simple universal function determined by the ratio of RG eigenvalues for the two relevant couplings at the Ising fixed point (these couplings are essentially $\beta_V - \beta_{c,Ising}$ and $1/\beta_T$), and $\lambda_{ren}$ is a parameter labelling the various curves. Then any observable $\mathcal{O}$ is a universal function of $\lambda_{ren}$ (modulo one or two nonuniversal scale factors), and we can label the possible continuum limits by their values of $\lambda_{ren}$:

(a) $\lambda_{ren} = -\infty$ corresponds to the $RP^{N-1}$ continuum theory: it is controlled by the "$RP^{N-1}$ fixed point", which is $Z_2$-gauge-invariant and has one marginally unstable direction (basically $1/\beta_T$).

(b) $\lambda_{ren} = +\infty$ corresponds to the conventional continuum $\sigma$-model: it is controlled by the "$N$-vector fixed point", which has one marginally unstable direction (basically $1/\beta$).

(c) $-\infty < \lambda_{ren} < +\infty$ corresponds to the one-parameter family of intermediate theories: they are controlled by the "critical Ising fixed point", which has one unstable direction with exponent $\nu = 1$ (basically $\beta_V - \beta_{c,Ising}$) and one direction which is either weakly unstable or marginally unstable (basically $1/\beta_T$).

This picture is confirmed in the Migdal-Kadanoff approximation, which predicts $g(\beta_T) = e^{-\text{const}(N) \times \beta_T}$ [1]. It is also confirmed by the exact solution in the limit $N \to \infty$ with $\beta_V$ and $\widetilde{\beta}_T \equiv \beta_T/N$ fixed [1, 4]. A more exact analysis can presumably be obtained by studying the renormalization properties of the expansion in powers of $1/\beta_T$ [4] around the Ising critical point $\beta_V = \beta_{c,Ising}$; we have not yet succeeded in doing this, but we do note that for $\beta_V = \beta_{c,Ising}$ there are *new* infrared divergences beyond those in the $N$-vector model, confirming our belief that *new* continuum theories — different from those of the $RP^{N-1}$ and $N$-vector models — can be obtained by suitable limits $\beta_T \to \infty$, $\beta_V \to \beta_{c,Ising}$.

We remark that the curve $m_T/m_V = 2$, which is the threshold for the generation of isotensor bound states, is one of the "curves of constant physics", corresponding to some particular finite value $\lambda_{ren} = \lambda_*$. For $\lambda_{ren} > \lambda_*$, the mass ratio $m_T/m_V$ is identically 2, but the scattering matrix of the isovector particles is presumably a nontrivial function of $\lambda_{ren}$.

Our family of continuum theories has a very peculiar feature: at the two extreme ends — the $RP^{N-1}$ and $N$-vector models — the iso*tensor* correlation functions are identical, after a suitable change of length scale and redefinition of the coupling, at all orders in perturbation theory in $1/\beta$. Nevertheless, at the nonperturbative level it seems clear that the isotensor correlation functions *cannot* be identical in the two theories: the isotensor two-point function of the $N$-vector theory presumably has a cut beginning at mass $m_T = 2m_V$ (i.e. Euclidean momentum $p = im_T$), corresponding to the scattering states of two isovector particles; while that of the $RP^{N-1}$ theory presumably has a *pole* at $m_T$, corresponding to a true isotensor particle ("bound state of two confined isovector constituents"). Apparently, nonperturbative corrections of order $e^{-4\pi\beta/(N-2)}$ — which in the continuum theory correspond to corrections of order $(p/\Lambda_{\overline{MS}})^{-2} \times$ logs, and are thus small at large momenta — contribute so strongly at low momenta as to change completely the particle structure of the theory. If this is in fact the case, then the perturbative $\Lambda$-parameter formula [5, 6] gives the change of length scale needed to match the correlation functions of the $RP^{N-1}$ and $N$-vector models *at high momenta* (i.e. much less than the ultraviolet cutoff but much greater than $m_T$), but it says nothing about the behavior at small momenta and in particular about the particle masses. In other words, the conventional prediction [7] for the nonperturbative constant $m_T/\Lambda_{\overline{MS}}$ in the $RP^{N-1}$ theory — namely that it takes the same value as in the $N$-vector theory, viz. twice $m_V/\Lambda_{\overline{MS}}$ (which is now known exactly [8]) — is apparently *invalid*.

The same argument shows that the intermediate theories have a nontrivial dependence on $\lambda_{ren}$, even in the isotensor sector, at least for $\lambda_{ren} \leq \lambda_*$. Indeed, in any of the theories with an isotensor bound state $(0 < m_T/m_V < 2)$, the isotensor correlation function has presumably a pole at $m_T$ *and* a cut at $2m_V$; so two theories with different mass ratios $m_T/m_V$ cannot possibly be equivalent modulo a length rescaling. Almost certainly this nontrivial dependence on $\lambda_{ren}$ continues even in the regime $\lambda_{ren} \geq \lambda_*$.

We have carried out extensive Monte Carlo runs on the mixed isovector/isotensor model (1) for $N = 3$ and $N = 4$, on $L \times L$ lattices up to $L = 512$ (resp. $L = 128$), using a Wolff-type embedding algorithm [9]. For $N = 3$ we have traced out the entire band-shaped region $5 \lesssim \xi_{max} \equiv \max(\xi_V, \xi_T) \lesssim 100$–$200$: this involved runs at 558 different triplets $(\beta_V, \beta_T, L)$, using a total CPU time of about 7 years on high-speed RISC workstations. We measured the susceptibilities $\chi_V, \chi_T$ and the second-moment correlation lengths $\xi_V^{(2)}, \xi_T^{(2)}$, among other observables. Our lattice sizes typically satisfy $3 \lesssim L/\xi_{max} \lesssim 10$, and we have carried out an extensive finite-size-scaling analysis. Details of the static and dynamic behavior can be found in [1] and [10], respectively; here we summarize the highlights of the former.

We constructed finite-size-scaling (FSS) curves for the observables $\mathcal{O} = \xi_V^{(2)}, \xi_T^{(2)}, \chi_V, \chi_T$ by considering all data points $(\beta_V, \beta_T, L)$ for which we have data also at $(\beta_V, \beta_T, 2L)$, and plotting $\mathcal{O}(2L)/\mathcal{O}(L)$ versus $\xi_{max}(L)/L$. All points corresponding to the same continuum-limit theory should then lie on a single curve (modulo corrections to scaling); if our claims about the universality classes are correct, these universality classes can be parametrized by (for example) the ratio $\xi_V^{(2)}/\xi_T^{(2)}$. We thus predict

$$\frac{\mathcal{O}(2L)}{\mathcal{O}(L)} = F_{\mathcal{O}}\left(\xi_{max}(L)/L \,;\, \xi_V^{(2)}(L)/\xi_T^{(2)}(L)\right) + O(L^{-\omega}) . \quad (3)$$

For the isovector (resp. isotensor) observables, we have divided $(\xi_V^{(2)}/\xi_T^{(2)})$-space into 12 (resp. 9) "slices" in which the scaling function $F_{\mathcal{O}}$ is found empirically to be constant within error bars, and in each slice we have chosen a smooth fitting function. The full set of curves is shown in [1]. The quality of the fits is amazingly good, and the curves vary smoothly from one slice to the next. More strikingly, the shape and even the *sign* of the finite-size corrections varies radically as a function of $\xi_V^{(2)}/\xi_T^{(2)}$.

These results provide strong evidence for the scenario proposed above, in two ways: First, the fact that all the data points with the same value of $\xi_V^{(2)}/\xi_T^{(2)}$ fall quite accurately onto a single curve — and this for all four observables — is strong evidence that (a) we are in the scaling region, and (b) all the theories with the same value of $\xi_V^{(2)}/\xi_T^{(2)}$ belong to the same universality class. Conversely, the fact that the curves for different values of $\xi_V^{(2)}/\xi_T^{(2)}$ are *not* the same is strong evidence that theories with different values of $\xi_V^{(2)}/\xi_T^{(2)}$ belong to *different* universality classes.

For the universality class of the $N$-vector theory, our FSS curves for $\xi_V^{(2)}$ and $\chi_V$ are in close agreement with those predicted by Flyvbjerg and Larsen [11] from the $1/N$ expansion through order $1/N^2$.

The contours of constant $\xi_V^{(2)}/\xi_T^{(2)}$ are shown in Figure 2. These curves confirm beautifully the scenario of Figure 1: it is perfectly believable that all these curves are heading for the Ising critical point $\beta_T = +\infty$, $\beta_V \approx 0.44$. The ratio $\xi_V^{(2)}/\xi_T^{(2)}$ is essentially constant at its asymptotic value $\approx 3.44$ in the entire region to the right of the last curve shown; in this area the theory is fully "$N$-vector-like".

For the pure $RP^{N-1}$ models with $N = 3, 4, 6, 8$, our simulations up to correlation length $\xi_T \approx 225, 25, 10, 10$ do not reveal asymptotic scaling. Therefore, we are unable to make any reliable statement about the nonperturbative constant $m_T/\Lambda_{\overline{MS}}$.

Our data for the $RP^2$ (resp. $RP^3$) model are also consistent with a different scenario [12], in which there is a critical point at $\beta_T \approx 5.73$ (resp. 6.96), only 3% (resp. 5%) beyond our last run. But for theoretical reasons we believe that such a critical point is unlikely: if it existed, then highly plausible (though unproven) correlation inequalities [1] would imply the existence of a critical curve continuing to the $N$-vector axis ($\beta_T = 0$), in contradiction with the believed (though unproven) asymptotic freedom of these models [13]. Moreover, the critical-point scenario works less well when we try to trace the presumed critical curve into the $(\beta_V, \beta_T)$-plane and to estimate the critical exponents. We think it more plausible that the apparent critical point arises from a rapid rise of $\xi_T$ in a crossover region.

The theoretical scenario proposed in this Letter applies also to other $\sigma$-models, for example the $SU(N)$ chiral model with mixed fundamental/adjoint action [14]. For $N \leq 4$ we predict the same scenario as in Figure 1, with new universality classes controlled by the critical $Z_N$ fixed point. For $N > 4$ the situation gets more complicated, as the $Z_N$ models are then believed to have an intermediate massless (spin-wave) phase [15] (this is *proven* for large $N$ [16]). We conjecture that in this case the curves of constant physics end up on the $\beta_T = +\infty$ axis precisely in the intermediate-phase interval, with $m_A/m_F = 0$ running to the first (Kosterlitz-Thouless) transition point and $m_A/m_F = 2$ running to the second (freezing transition). Since the difference between $SU(N)$ and $SU(N)/Z_N$ is invisible in conventional perturbation theory, these scenarios are in conflict with the conventional wisdom.

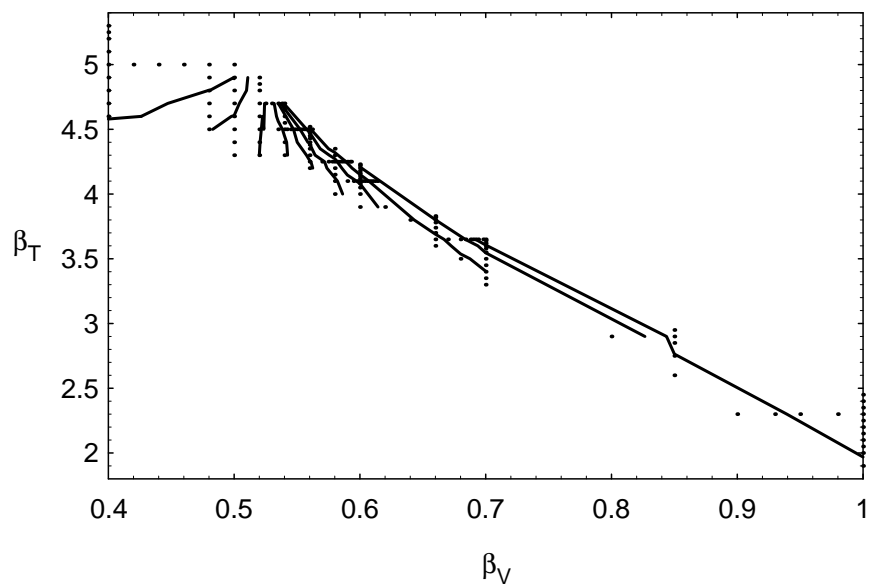

Figure 2: Contour lines of constant ratio $\xi_V^{(2)}/\xi_T^{(2)} = 0.3$, 0.5, 0.75, 1.0, 1.25, 1.5, 2.0, 2.5, 3.0, 3.2 in the $(\beta_V, \beta_T)$-plane: expanded view of the region $0.4 \leq \beta_V \leq 1.0$. Dots indicate our runs.

We predict a similar scenario for the *complex* mixed isovector/isotensor model, which interpolates between the $2N$-vector and $CP^{N-1}$ models. In this case we conjecture that the curves of constant physics end up on the $\beta_T = +\infty$ axis between $\beta_V = \beta_{c,XY}$ and $\beta_V = +\infty$; this picture is consistent with the large-$N$ calculations of Campostrini and Rossi [17]. Here, however, the manifolds $S^{2N-1}$ and $CP^{N-1} \simeq S^{2N-1}/U(1)$ *are* different at the perturbative level; so our prediction, though apparently new, does not contradict the conventional wisdom.

Similar scenarios can be expected for lattice gauge theories with mixed fundamental/adjoint action [18] whenever the gauge theory based on the center of the original gauge group has a *second-order* (critical) transition.

The results of this Letter show that the complete information about a continuum quantum field theory cannot be deduced solely from its formal continuum action. Rather, it is necessary to impose an ultraviolet cutoff (e.g. choose a lattice action), search the phase diagram of the cutoff theory for critical points, and study the possible continuum limits. We suspect that there are many hitherto-unexpected continuum quantum field theories waiting to be discovered.

We wish to thank M. Aizenman, P. Butera, M. Campostrini, M. Caselle, M. Creutz, C. Destri, F. Gliozzi, H. Kunz, M. Lüscher, F. Niedermayer, G. Parisi, P. Pasini, D. Rokhsar, P. Rossi, E. Seiler, T. Spencer, E. Vicari, U. Wolff, G. Zumbach and D. Zwanziger for helpful discussions. Computations were performed at SCRI, NYU, SNS, the Università di Pisa and the Pittsburgh Supercomputer Center. The authors' research was supported in part by the INFN and CNR, DOE contracts DE-FC05-85ER250000 and DE-FG02-90ER40581, NSF grants DMS-8911273 and DMS-9200719, NATO grant CRG 910251, and a grant from the NYU Research Challenge Fund.